# Does double-blind peer review reduce bias? Evidence from a top computer science conference

Mengyi Sun[1], Jainabou Barry Danfa[2], and Misha Teplitskiy[2,3*]

**Affiliations**

[1]Department of Ecology and Evolutionary Biology, University of Michigan, Ann Arbor, MI, 48109

[2]School of Information, University of Michigan, Ann Arbor, MI, 48104

[3]Laboratory for Innovation Science at Harvard, Boston, MA, 02134

**Abstract**

Peer review is widely regarded as essential for advancing scientific research. However, reviewers may be biased by authors' prestige or other characteristics. Double-blind peer review, in which the authors' identities are masked from the reviewers, has been proposed as a way to reduce reviewer bias. Although intuitive, evidence for the effectiveness of double-blind peer review in reducing bias is limited and mixed. Here, we examine the effects of double-blind peer review on prestige bias by analyzing the peer review files of 5027 papers submitted to the International Conference on Learning Representations (ICLR), a top computer science conference that changed its reviewing policy from single-blind peer review to double-blind peer review in 2018. We find that after switching to double-blind review, the scores given to the most prestigious authors significantly decreased. However, because many of these papers were above the threshold for acceptance, the change did not affect paper acceptance decisions significantly. Nevertheless, we show that double-blind peer review may have improved the quality of the selections by limiting other (non-author-prestige) biases. Specifically, papers rejected in the single-blind format are cited more than those rejected under the double-blind format, suggesting that double-blind review better identifies poorer quality papers. Interestingly, an apparently unrelated change – the change of rating scale from 10 to 4 points – likely reduced prestige bias significantly, to an extent that affected papers' acceptance. These results provide some support for the effectiveness of double-blind review in reducing prestige bias, while opening new research directions on the impact of peer review formats.

* To whom correspondence should be addressed: Misha Teplitskiy, tepl@umich.edu

# Introduction

The role of peer review in the advancement of scholarly knowledge is hard to overstate (*1*). The stakes are high: a single misjudgment can kill a promising project, ruin a budding research career, or even delay a life-saving breakthrough. Ideally, peer review should be fair: the evaluation of scientific work should not be based on anything other than the work itself. In reality, peer review, like other human judgments, may be subject to a number of biases (*2*). Evidence for biased evaluation of scientific work based on factors such as author prestige (*3*) and even demographic characteristics like gender (*4*) or race (*5*) is prevalent. Even when the reviewers are unbiased, their judgments may be highly unreliable: an accurate assessment of new scientific work is notoriously difficult (*6*). Given the importance of peer review, policies that reduce the bias and increase the reliability of the peer review process are needed. Here, we consider the policy of double-blind review: masking authors' identities to reviewers.

The form of peer review used by many journals is single-blind, in which the reviewers are unknown to the author, but the identities of the authors are known to the reviewer (*7*). Reviewers may infer from authors' identities, and in particular prestige, a number of factors that affect how they review a paper. Therefore, an intuitive solution to reduce reviewing bias is to adopt the so-called double-blind peer review (*7*), in which both the authors and the reviewers are anonymous. This idea, despite its appealing logic, is not foolproof: no reviewing system, including double-blind peer review, can achieve perfect anonymity. Sometimes the submitted work itself provides enough information for guessing the identities of the authors (*8, 9*). Moreover, most of the conferences or journals that adopt double-blind peer review do not prohibit the authors from posting their work on preprint servers before submission for the purpose of accelerating scientific communications. This practice potentially results in deanonymization and hence dilutes or negates the effort of double-blind peer review (*10*).

Because of the above concerns, whether double-blind peer review is effective in reducing bias in practice is unclear. Past studies on the efficacy of double-blind peer review generally demonstrated positive effects, with some heterogeneity. For example, Okike et al. (*11*) devised an experiment to show the effect of double-blind peer review in reducing prestige bias. In the experiment, they fabricated a manuscript for which two past presidents of the American Academy of Orthopedic Surgeons were listed as authors. With the help of an orthopedics journal, the manuscript was sent out for review to 119 reviewers, randomly assigned to be under double-blind or single-blind conditions. They found that the manuscript was more likely to be accepted in the single-blind setting. Recently, Tomkins et al. (*12*) designed a similar experiment where papers submitted to a reputable computer science conference were subjected to both single-blind and double-blind peer review. They found that papers authored by famous authors and (or) authors from prestigious institutions were rated higher and more likely to be recommended for acceptance in the single-blind setting. In contrast to the above results, a study by Fisher et al. (*13*) found that works from more productive authors actually were evaluated higher in the double-blind setting. Other experiments showed more complex patterns. For instance, in an experiment conducted by Blank (*14*), 1498 papers submitted to The American Economic Review were randomly assigned to single-blind or double-blind peer review. Interestingly, for authors from top- or bottom-rank institutions, no significant differences in acceptance rates were observed between double-blind peer review versus single-blind peer review (*14*). However, authors from mid-tier institutions benefited from single-blind peer review. Blank also found that female authors performed slightly better under double-blind peer review, although the effect is not significant. Paradoxically, although it was assumed that authors from foreign countries were biased against, they performed better under single-blind peer review (*14*). Thus, the available evidence shows that the effects may be inconsistent and non-linear across prestige. Furthermore, none of the above studies directly addressed the issue of whether double-blind peer review improves the reviewing quality in the sense of better distinguishing between high-quality research and low-quality research.

Here, we analyze peer review data from the International Conference on Learning Representations ICLR) (see Materials and Methods for details). ICLR is a highly prestigious conference in machine learning (*15*). Prior to the year 2018, ICLR used single-blind peer review. Starting from the year 2018, ICLR switched to double-blind peer review. Most importantly, the peer review file for each paper submitted to ICLR can be obtained through OpenReview (*16*), a web-based platform that aims to facilitate the free dissemination of peer review activities. The peer review data associated with the sudden policy change provide a unique opportunity to address the efficacy of double-blind peer review in reducing reviewer bias and improving the quality of the review. Author prestige was measured by the percentile of the mean author citations up to the year of submission obtain from the Microsoft Academic Graph (MAG) (*15*) database (See materials and Methods). The quality of decisions was measured by the citations the papers received after acceptance or rejection.

## Results
**The non-linear effects of double-blind peer review on prestige bias**
To test the hypothesis that double-blind peer review can reduce author prestige bias, we first estimated a linear regression model with the specification:

$$Mean\_score = Prestige + double\_blind \times Prestige + is\_2018 + is\_2019 + \varepsilon.$$

Where *Mean_score* is mean-rating among the three reviewers of a paper and $\varepsilon$ is the error term. In this regression, we focused on papers submitted from 2017 to 2019 because the rating scales are the same in these three years, whereas in 2020, a different rating scale was used. Note that *double_blind* is used only as an interaction term because its main effect is captured by *is_2018* and *is_2019*. As shown in Table 1, the estimated coefficient of the interaction effect was not statistically different from 0 ($\beta$=0.096, $p$=0.68), suggesting that scoring across prestige levels did not change significantly across reviewing formats. The result for the interaction effect is similar if instead of mean score we use paper acceptance as the outcome variable (Table 2, $\beta$=0.034, $p$= 0.68).

[ Table 1 about here ]

[ Table 2 about here ]

At face value, this seems to suggest that double-blinding does not have a significant effect in reducing author prestige bias. However, linear regression assumes linearity and might not be able to detect a nonlinear effect in small samples, whereas previous research suggests the bias might not be linear with respect to prestige. Consequently, we turn to nonparametric techniques. We divided the papers in each year into three equally sized sets based on mean author prestige. For each of these groups, we compared mean scores in 2017 (single-blind) and 2018 (double-blind). As shown in **Fig.1A**, the mean-ratings of papers with the highest one-third prestige in the double-blind setting is significantly lower than the papers with similar prestige in the single-blind setting ($p$=0.0035, Mann–Whitney U test). However, for papers with median or low prestige, no statistically significant difference can be observed between these two settings. As a negative control, we performed the same comparisons for papers submitted in 2018 versus 2019, where both years used double-blind peer review. As shown in **Fig.1B**, none of the comparisons of mean-ratings between papers with similar prestige are significant. These results support the hypothesis that double-blind peer review reduces the prestige bias. However, the effect is non-linear: on the one hand, double-blinding reduces the "premium" that top authors can gain from their reputation; on the other hand, it does not significantly boost the low-reputation authors in terms of raw evaluations, at least in our data.

[ Figure 1 about here ]

Next, we asked whether the apparent reduction in the prestige premium in ratings under double-blinding will be translated into differences in acceptance. This is not a trivial question because whether reducing bias in rating results in reducing bias in paper acceptance depends on whether the affected manuscripts are around the acceptance threshold. If the affected papers are sufficiently above the threshold even after removing the premium, then ratings may be debiased but final decisions materially unaffected. In 2017 and 2018, the acceptance rate is relatively high: ~50% under our definition of accepted papers (see Materials and Methods for our definition of acceptance). Given that double-blinding only affected papers with the highest reputations but not the other groups, it is plausible that the difference in mean-ratings will not translate into a difference in acceptance rates. As expected, this is indeed the case (**Fig.1C**).

Other than the effect on mean-rating, are there any other effects that double-blinding can have on the reviewing process? We hypothesized that because double-blinding masks the authors' identities, reviewers will have fewer clear and agreed-upon cues to use as a heuristic during the reviewing process. Lacking such cues and needing to base their judgment on the more difficult to parse and contested manuscript contents, we expect that reviewers will disagree more in their judgments. To test this hypothesis, we calculated the standard deviations (*SD*) among the three reviewers for each paper. We then compare the *SD* values under double-blind peer review (2018) versus single-blind peer review (2017). Indeed, the standard deviations are generally larger in the double-blind setting than the single-blind setting (**Fig.1D**). As a negative control, we compared the *SD* values in the year 2018 and the year 2019. No significant differences in sd values are observed in any reputation group (**Fig.1E**). Unlike the effect of double-blinding on mean rating, the effect of double-blinding on *SD* seems to affect a wider range of papers. For example, it affects both the high-reputation group and the median-reputation group significantly. For the low-reputation group, the p-value is small but not significant at conventional levels ($p$=0.079, Mann–Whitney U test), indicating that the effect might exist even for the low-reputation group but might need a larger sample size to detect. These results suggest that double-blinding increases the disagreement between reviewers by masking easily observable and non-controversial cues. Interestingly, although an increased level of reviewer-disagreement usually implies a lower reviewing quality(*17*) because of the reduction of precision, this is not necessarily so in our case. More specifically, if the reduction of precision is due to the removal of bias, the reviewing quality can actually improve. We illustrate this point in the next section.

**Double-blind peer review more effectively identifies low-quality papers**
A less biased reviewing process should better differentiate low-quality papers from high-quality papers. The results above suggest that double-blind review is indeed less biased by author prestige. However, author prestige is just one of a number of other biases that may be present, such as affiliation prestige or biases against particular demographic groups. A natural way to test for the presence of such biases together is by the quality of ultimate selections across reviewing formats, which we address here.

In 2017 and 2018, ICLR has similar acceptance rates ($p=0.50$, Fisher's exact test). Furthermore, when comparing the distributions of author prestige among papers in 2017 versus 2018, we found that the distributions are similar ($p=0.13$, Kolmogorov–Smirnov test). Overall, we expect the quality of papers in the submissions pools to be of similar quality as well. These similarities make comparing the efficiency in differentiating low-quality papers from high-quality papers in 2017 (single-blind) versus 2018 (double-blind) an ideal test for our hypothesis that double-blinding increases the accuracy of peer review.

Specifically, we predicted that (1): for accepted papers, papers accepted in ICLR 2017 should have fewer citations compared to papers accepted in ICLR 2018 in the same time window, and (2): for rejected papers, papers rejected in ICLR 2017 should have more citations compared to papers rejected in ICLR 2018. Ideally, we would like to test both (1) and (2). However, once a paper is accepted by a prestigious conference such as ICLR, it is likely to receive a boost in citations(*18*). If a paper is published only after acceptance, it enjoys this boost at the very beginning of the publication, whereas accepted papers published (e.g., on Arxiv) before acceptance will not enjoy this boost from the start. Therefore, the citations garnered by an accepted paper in a fixed time window depends on its quality and the posted time relative to the decision date. ICLR does not prevent authors from posting online before submission, even in the years of double-blind peer review, so the posted time is heterogeneous in each year. If the distributions of posted time relative to the decision date are similar across years, the comparison is still possible. However, in our dataset, we observed a substantial behavior shift in posting papers when ICLR switch to double-blind peer review: in 2017, only a small fraction of papers (12.3%) were posted after the decision date, whereas in 2018, more than half of the papers (54.9%) were posted after the decision date[a]. This behavior shift makes it difficult to test prediction (1).

Therefore, we focused our analysis on prediction (2).We collected the 2-year-citation of each rejected paper in ICLR 2017 and ICLR 2018 from the MAG database (see Materials and Methods) and compare their median citation level. As shown in Fig.2, the citations of rejected papers in ICLR 2018 is indeed significantly lower than ICLR 2017 ($p=0.0016$, Mann–Whitney U test), despite similar rating percentiles within papers published in the same year ($p=0.14$, Mann–Whitney U test). We further reasoned that, because papers posted after the decision date in ICLR should include a larger fraction of papers that are successfully anonymized (since posted before the decision date increases the risk of de-anonymization (*10*), these papers should be rejected more accurately. And **Fig.2** shows, this is indeed the case ($p=4.4\times10^{-11}$, Mann–Whitney U test). Notice that this is not because the (rejected) papers posted after the decision rate have lower subjective quality: the percentile rating is similar between rejected papers in ICLR 2017 and rejected papers in ICLR 2018 posted after the decision date ($p=0.60$, Mann–Whitney U test). Thus, our results support the hypothesis that double-blinding improves reviewing accuracy.

---

[a] This pattern may be viewed optimistically: submitters were acting in good faith to improve the efficiency of double-blinding.

[ Figure 2 about here ]

**Rating-scale change might unexpectedly reduce prestige bias**
The above results suggest that double-blinding does reduce reputation bias. However, the effect is subtle and might not be strong enough to affect the decision. Furthermore, given that double-blinding is never perfect and sometimes impractical (such as NIH grant review), it is worth asking whether there are other mechanisms that have a significant effect on reducing prestige bias. Drawing on a previous randomized experiment which showed that changing rating scales from a fine-grained rating scale to a more coarse rating scale could reduce gender bias in evaluations (*19*), we consider the effects of rating scale. Two potential mechanisms might explain why a coarse rating scale could reduce bias. First, a coarser scale might remove from reviewers a way to express subtle, and seemingly innocuous, preferences for more reputable authors (*19*). Second, reviewers might perceive the highest score for a coarser scale differently than the highest scores for a finer scale, for example by believing that only exceptional talents, evidenced by previous accomplishments, can achieve such high scores (*19*). In our case, both mechanisms might be at work: in 2020, ICLR changes the scale from a 10-categories rating scale (rating 1-10) to a 4-categories rating scale (1,3,6,8), presumably to reduce reviewer cognitive burden. Here, we examine whether this change affected acceptance of prestigious authors.

Because ICLR 2019 and ICLR 2020 have similar acceptance rates ($p$=0.24, Fisher exact test) and similar paper input in terms of author prestige ($p$=0.098, Kolmogorov–Smirnov test), a comparison between the outcomes of paper submissions in ICLR 2019 versus ICLR 2020 might shed light on whether scale-changes can potentially reduce reputation bias. To this end, we perform panel regression with the outcome variable as paper acceptance, whereas the key independent variable is the interaction between scale-change and paper prestige. Here we did not perform a panel regression with respect to mean-rating because, after the rating scale change, the mean-ratings are no-longer comparable in ICLR 2019 and ICLR 2020, and the interpretation of results will be unclear. Estimated coefficients are displayed in **Table 3**. The rating scale change shows a significant effect in reducing prestige bias ($\beta$= 0.12, $p$=0.029). More specifically, before the rating-scale change, a reduction of prestige by ten percentiles (e.g., from top 1% to top 11%) was associated with a reduction of acceptance rate by about 3 percent on average. After the rating-scale change, a reduction of prestige by ten percentiles is associated with a reduction of only 1.7 percent on average. This result supports our hypothesis that scale-change can reduce reviewer bias. Given the recency of the data, sufficient citation trajectories are unavailable and it remains to be seen whether bias-reduction by changing to a more coarse rating scale will improve reviewing accuracy.

[ Table 3 about here ]

## Discussion
The importance of understanding how peer review format affects bias in science is tremendous, but existing studies are few and often generate mixed results. Here, we contribute to the literature on bias in peer review by utilizing a sudden policy change in the reviewing format of ICLR, a top computer science conference. Overall, we found that relative to single-blind reviewing, double-blinding reduces prestige bias. However, double-blinding did not affect authors with different reputations uniformly: double-blinding reduced the scores gives to submissions from the top third of authors but did not provide a significant boost in scores to authors with low or medium prestige. Furthermore, the nonlinear effects of double-blinding on scoring did not significantly affect the ultimate acceptance outcomes, likely because the most affected papers (from top authors) were above the bar for acceptance even without the premium. Nevertheless, double-blinding improved the efficiency in rejecting papers: papers rejected under the presumably more meritocratic double-blind format were of lower quality (as measured by citations) than those rejected under-single blind format. Finally, we showed that measures other than double-blinding, such as changing the rating scale, might also appreciably reduce reviewer bias.

This work is not without limitations. First, although the effect of double-blinding is certainly heterogeneous among author reputation groups, it does not wipe out the possibility that double-blinding also impacts the rating of authors with lower reputations. It is possible that, with larger samples and better author disambiguation, the effect of double-blinding on authors with low reputations can be revealed. Second, newer papers usually accumulate citations faster than older papers due to the yearly growth of scientific activities (*20*), which might potentially render the comparison between papers published in different years not directly comparable. However, this time trend would render the 2-year-citations of rejected papers in the double-blind setting (year 2018) higher than the single-blind setting (year 2017), which is in the opposite direction of our observations. Therefore, our results should be conservative. Third, and perhaps most importantly, we used 2-year-citations as a proxy of paper quality. This is by no means a perfect measure of paper quality. Specifically, the correlation between short-term citations and long term impact is noisy (*6*). It remains to be seen whether our results also hold for long term impact. Furthermore, novel papers tend to suffer from delayed recognition and often have lower impacts even in the long-term (*21*). So an alternative explanation of our results will be double-blind peer review is harsher to novel work, which might be because highly reputable authors are no longer able to "sell" novel yet risky ideas with the help of their reputation (*22*). If this is the case, and if our purpose is to encourage novel work, double-blind peer review might not be ideal. Hence, it is important to verify whether novel papers are harder to publish under double-blind peer review when more meta-data for each paper is available (*21*). Most importantly, the effects of a policy intervention like peer review change are ideally studied with randomized controlled trials. Here, we rely on a presumably exogenous policy change, but we cannot rule out that the change affected submission behavior after it was made, making the pool of submissions and/or reviewers systematically different before and after the change.

Besides providing supportive evidence for the benefit of double-blinding, our work also suggests several new research directions. First, under what conditions does reducing bias in *ratings* minimize bias in *acceptance*? As shown in our results, we did not detect a significant reduction of reputation bias in paper acceptance despite finding a reduction in rating bias. Although this may be because the sample size of our study is too small, the problem of whether reducing rating bias will affect results will likely persist. To put this at the extreme, if the acceptance rate is 100% or 0%, then rating bias won't matter. In general, there will be a range of acceptance rates that will translate the impact of bias reduction in rating maximally. Exploring this question using mathematical modeling when we have a better sense of the functional form of bias will be important to setting optimal acceptance rates. Second, is the quality of accepted papers higher in the double-blind setting? As mentioned in our results, we cannot answer this question conclusively due to the limited sample size and recency of the data. Unlike short term citations, long-term citations are insensitive to the technical difficulties we mentioned (*6*). Therefore, this question can be addressed while a much longer time window, e.g. 10 years, is used (*6*). Lastly, further research is needed to understand the effects of rating scales on reducing bias. In particular, what are the cognitive mechanisms behind scale-change effects, and do they affect reviewing accuracy? Answering these questions will be important in designing alternative reviewing schemes that reduce bias and increase reviewing quality, especially when double-blinding is impractical, such as NIH grant review.

## Materials and Methods

### ICLR paper data

We obtained the information of papers submitted to ICLR from 2017 to 2020 using the OpenReview(*16*) python API: https://openreview-py.readthedocs.io/en/latest/index.html. Information collected includes the title of each paper, the names and email address of each author, the year of submission, the reviewer ratings, and the final decisions. From 2017 to 2019, ICLR adopted a 10-point rating scale (1-10), whereas, in 2020, ICLR adopted a 4-point rating scale(1, 3, 6, and 8). For the year 2017-2018, the decision categories are: "Accepted (oral)", "Accepted (poster)", "Invited to workshop", and "Reject". For year 2019, the decision categories are: "Accepted (oral)", "Accepted (poster)", and "Reject". For year 2020, the decision categories are: "Accepted (Talk)", "Accepted (spotlight)", "Accepted (poster)", and "Reject".

### Calculating prestige of each paper

To calculate the prestige of each paper, we preprocessed the ICLR paper data to get the information of 12694 authors from all papers. We then obtained the unique identifiers (mag id) of all authors from the Microsoft Academic Graph (MAG) database using the MAG REST API("evaluate" and "interpret" methods):

https://docs.microsoft.com/en-us/academic-services/project-academic-knowledge/reference-evaluate-method, and

https://docs.microsoft.com/en-us/academic-services/project-academic-knowledge/reference-interpret-method.

To increase the accuracy of matching, for each author, we extracted his/her institution from their email domain name when possible (some email addresses, such as 'Gmail', are not informational). In the query, we attempted to limit the query to the particular institution and the field "Computer Science". If a match can not be found, we relax the criteria by removing either institution or field or both institution and field in our query. After the matching process, we manually examine a random subsample of 100 matched authors (Table S1) and found the matching accuracy to be reasonable (78%).

Next, for each author, we download the citation history of each paper authored by that author from the MAG database. This allows us to compute the total citations of an author up to a given year. For each paper, we calculated the average citations up to the year of submission of all authors that we can find a match in the MAG database. The above procedure allows us to get the prestige measure for 5027 papers.

### Linear regressions of mean-rating or acceptance on prestige

To test whether double-blind has an effect in reducing prestige bias, we ran the following OLS regressions on data from 2017 to 2019:

$$Mean\_score = Prestige + double\_blind \times Prestige + is\_2018 + is\_2019 + \varepsilon, \text{ and}$$
$$Acceptance = Prestige + double\_blind \times Prestige + is\_2018 + is\_2019 + \varepsilon \,.$$

The *Mean_score* is the mean ratings among the three reviewers for each paper. The *Acceptance* variable is a binary variable. It takes the value 1 when the paper is not rejected. Otherwise, it takes the value 0. This definition of *Acceptance* allows us to achieve a more balanced number of accepted papers versus rejected papers, increasing our statistical power. The *is_2018* and *is_2019* are binary variables specifying whether or not the paper is submitted in a given year. These variables are used to control potential time trends in the data. Noted that the *double_blind* variable is basically a year variable (take the value 0 if the paper is submitted to ICLR 2017. Otherwise, it is 1, so the main effect of *double_blind* will be absorbed into the year variables. The 'Prestige' for a paper is the percentile of its mean author citations among all papers submitted to ICLR in the same year. In this case, we rank the paper prestige in descending order, so the lower the value of this variable, the higher the prestige. Finally, the $\varepsilon$ is the error term.

We ran a similar regression to test the effect of scale-change, using paper data from ICLR 2019 and ICLR 2020:

$$Acceptance = Prestige + Rating\_change \times Prestige + is\_2020 + \varepsilon.$$

The *Rating_change* variable is a binary variable specifying whether the paper is submitted to ICLR 2020, which adopted a new rating scale. And therefore, its main effect is captured in the variable *is_2020*.

We calculate the *p*-values using robust standard errors, and the results are presented in Table 1-3.

**Citation data for papers submitted to ICLR2017 and ICLR2018**

We matched papers submitted to ICLR 2017 and ICLR 2018 to the MAG database using the paper title and limited the query to the field of "Computer Science". For each matched paper, we obtained the publication date from the MAG database. If a paper appears in multiple venues (posted on arxiv before submission, for example), we define the earliest publication date among all venues as its publication date. We then remove papers published before Jan 1, 2016, and papers published after Sep 30, 2018, in order to remove mismatched papers and allow at least two years for accumulating citations. The vast majority of matched papers are retained (1368 out of 1400). Finally, for each paper, we computed the citations accumulated within two years of its publication date from all venues.

# Tables and Figures

**Table 1. Association of double-blind peer review and prestige with mean-score**

|  | β | SE | t | p |
|---|---|---|---|---|
| *(Intercept)* | 6.141393 | 0.120959 | 50.7723 | $<2.2\times10^{-16}$ |
| *is_2018* | -0.299507 | 0.133706 | -2.2400 | 0.02517 |
| *is_2019* | -0.291256 | 0.131796 | -2.2099 | 0.02719 |
| *prestige* | -0.916086 | 0.214493 | -4.2709 | $2.011\times10^{-5}$ |
| *double_blind x prestige* | 0.096581 | 0.230967 | 0.4183 | 0.67573 |

$R^2=0.04389$, n=2814

**Table 2. Association of double-blind peer review and prestige with acceptance (linear probability model, accept=1)**

|  | β | SE | t | p |
|---|---|---|---|---|
| *(Intercept)* | 0.663172 | 0.043927 | 15.0973 | $<2.2\times10^{-16}$ |
| *is_2018* | -0.051001 | 0.049926 | -1.0215 | 0.3070895 |
| *is_2019* | -0.164830 | 0.049088 | -3.3579 | 0.0007959 |
| *prestige* | -0.323637 | 0.075412 | -4.2916 | $1.834\times10^{-5}$ |
| *double_blind x prestige* | 0.033872 | 0.082490 | 0.4106 | 0.6813844 |

$R^2=0.04545$, n=2814

**Table 3. Association of rating scale change and prestige with acceptance (linear probability model)**

|  | β | SE | t | p |
|---|---|---|---|---|
| *(Intercept)* | 0.501863 | 0.025755 | 19.4862 | $<2.2\times10^{-16}$ |
| *is_2020* | -0.103623 | 0.032663 | -3.1725 | 0.001524 |
| *prestige* | -0.296802 | 0.041679 | -7.1211 | $1.286\times10^{-12}$ |
| *is_2020 x prestige* | 0.121279 | 0.053279 | -2.2763 | 0.02887 |

$R^2=0.0213$, n=3624

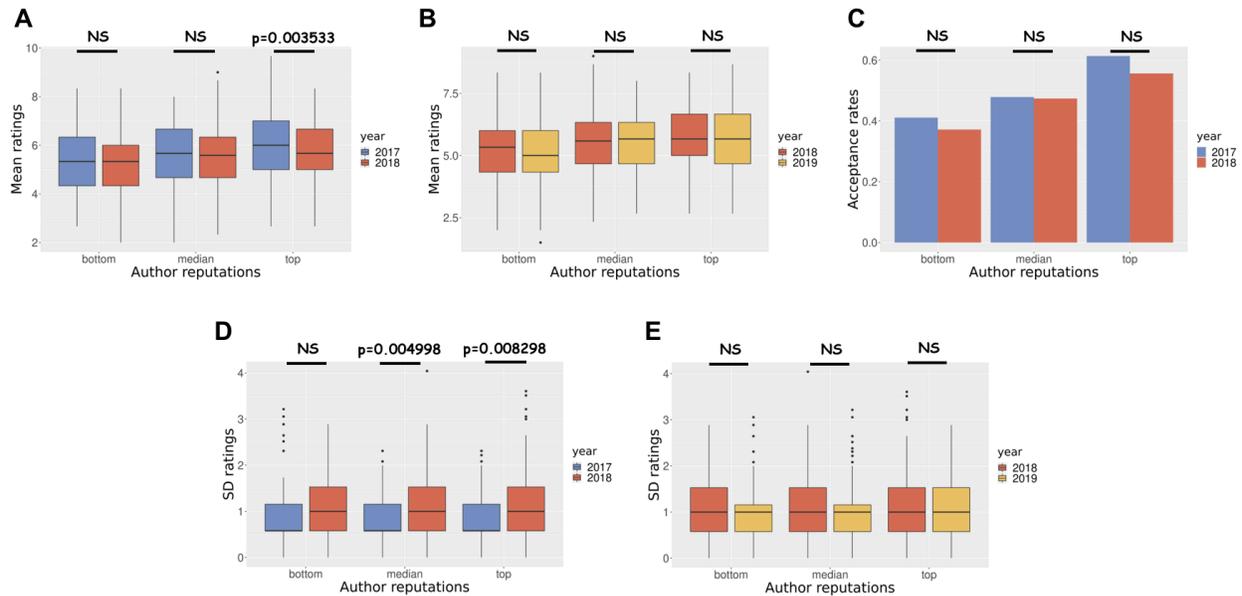

**Fig.1. The effect of double-blinding on reviewer rating and paper acceptance.** (A-B) Double-blind peer review reduced the mean reviewer rating for top authors but not the other authors(the year 2017: single-blind; the year 2018 and 2019: double-blind. NS: not significant). (C) Double-blind peer review did not necessarily result in differences in acceptance rates. (D-E) Double-blind peer review increases the disagreement (measured by standard deviation (*SD*)) among reviewers.

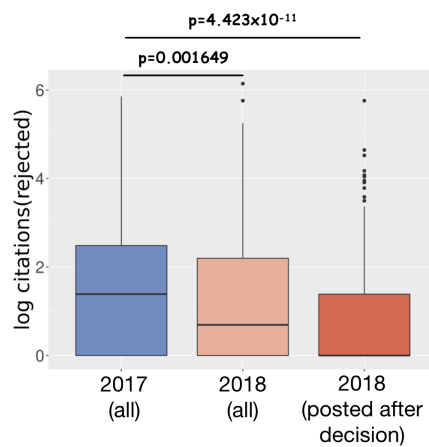

**Fig.2. The effect of double-blinding on reviewing quality.** Papers rejected in the double-blind setting (the year 2018) garnered significantly lower 2-year-citations compare to papers rejected in the single-blind setting (the year 2017), indicating lower quality. The effect is especially prominent when considering papers published after the decision date, which were more effectively anonymized.